\documentclass[twocolumn,showpacs,amsmath,amssymb]{revtex4}
\usepackage{graphicx}

\begin{document}

\title{Topology and Geometric Structure of Branching MERA Network}
\author{Hiroaki Matsueda\footnote{matsueda@sendai-nct.ac.jp}}
\affiliation{Sendai National College of Technology, Sendai 989-3128, Japan}
\date{\today}
\begin{abstract}
We examine a bulk-edge correspondence of branching MERA networks at finite temperatures in terms of algebraic and differential topology. By using homeomorphic mapping, we derive that the networks are nonorientable manifolds such as a M$\ddot{\rm o}$bius strip and a Klein bottle. We also examine the stability of the branch in connection with the second law of black hole thermodynamics. Then, we prove that the MERA network for one-dimensional quantum critical systems spontaneously separates into multiple branches in the IR region of the network. On the other hand, the branch does not occur in more than two dimensions. The result illustrates dimensionality dependence of spin-charge separation / coupling. We point out a role of twist of the surfaces on the phase string between spinon and holon excitations.
\end{abstract}
\maketitle

\section{Introduction}

\subsection{Information of Quantum Criticality Incorporated into MERA Networks}

The multiscale entanglement renormalization ansatz (MERA) is variational one appropriate for quantum critical systems~\cite{Vidal}. The MERA for spatially $d$-dimensional ($d$D) systems is represented by a product of tensors, but the tensor network extends in a $(d+1)$-dimensional 'holographic' space with a discrete AdS${}_{d+2}$ metric. The object propagating in this extra dimension is quantum entanglement. A measure of the magnitude of the entanglement is called entanglement entropy, and usually the entropy obeys the well-known area law scaling~\cite{Bekenstein,Hawking,Bombelli,Srednicki}. In 1D critical systems, the area law scaling is logarithmically violated~\cite{Holzhey,Calabrese}, and the MERA represents the violation in terms of differential geometry in the holographic space~\cite{Vidal,Ryu}.

In general, non-locality or entanglement of quantum states can not be characterized by local order parameters. However, even in such a case the entanglement entropy is still a good probe or a kind of the order parameter that can detect essential features of the quantum states as they are. Thus, the presence of the proper entanglement-entropy scaling within the MERA network for the critical systems leads to strong constraint to the electronic properties described by the network. This means that we may do it without handling a specific Hamiltonian. For instance, in the Kondo problem, we can determine the Kondo temperature from the wave function directly by considering a condition that the gain of the entanglement between the impurity and conduction electrons becomes stronger than the loss of the entanglement arising from temperature fluctuation~\cite{Matsueda}. The reason for such powerfulness is supported by recent observations that the tensor network contains enough information of conformal field theory (CFT)~\cite{Cirac,Poilblanc,Lou,Tanaka,Matsueda2}.

\subsection{Motivation and Purpose}

In this paper, we discuss whether the MERA network can distinguish the spin-charge separation with the spin-charge coupling, depending on spatial dimensionality of our target models. Since they are not classical orders, we may do it with the help of the entanglement entropy. Although we do not calculate physical quantities by using a specific Hamiltonian, our attention is of course the $t$-$J$ model defined by
\begin{eqnarray}
H= -t\sum_{\left<i,j\right>,\sigma}\left(\xi_{i,\sigma}^{\dagger}\xi_{j,\sigma}+\xi^{\dagger}_{j,\sigma}\xi_{i,\sigma}\right) + J\sum_{\left<i,j\right>}\vec{S}_{i}\cdot\vec{S}_{j} ,
\end{eqnarray}
where $\xi_{i,\sigma}^{\dagger}=c_{i,\sigma}^{\dagger}(1-n_{i,-\sigma})$, $\xi_{i,\sigma}=c_{i,\sigma}(1-n_{i,-\sigma})$ are creation and annihilation operators of and electron with spin $\sigma$, at site $i$, $\vec{S}_{i}=(1/2)\sum_{\sigma,\sigma^{\prime}}c^{\dagger}_{i,\sigma}(\vec{\sigma})_{\sigma\sigma^{\prime}}\xi_{i,\sigma^{\prime}}$, and $t$ and $J$ are electron hopping and exchange coupling.

In spatially 1D cases, the $t$-$J$ model (or the large-$U$ Hubbard model) has been solved by various theoretical and numerical approaches~\cite{Ogata,Sorella,Penc,Penc2,Nakamura}. The most direct one is the Bethe ansatz solution. There, an electron is separated into the spin and the charge collective excitations, and they form edge singularities in the electron's one-particle spectral function. In viewpoints of bosonization, these excitations are typical examples of two-components Tomonaga-Luttinger liquids, and each of them can be independently described by the CFT with the central charge $c=1$.

In that sense, we may think that the original MERA network is oversimplified to describe multiple-compoments CFTs, although it would be the best one for single-component CFTs. Then, it is natural to consider the network spontaneously separates into two independent AdS-like subspaces in IR region. Such a concept has been introduced recently, and is called 'branching' MERA network~\cite{Evenbly}. Unfortunatelly, the method for holographic bifurcation introduced previously is rather technical. Thus, we give more physical meaning to the branching network.

For this purpose, we investigate algebraic and differential topology inherent in the network at finite temperatures. Up to now, our analysis was based on a discrete version of differential geometry, since the entanglement entropy in the holographic space is proportional to a geodesic distance connecting two end points of a partial system with length $L$ on the original 1D lattice. Apart from such a picture, now we introduce a topological view for the MERA network, which requires to look at global structure of the network independent of the local metric. As will be precisely defined in the main part of this paper, the network at finite temperatures can be identified to be a topological manifold, and is characterized by the number of genera created in the IR region of the network by the presence of multiple electronic excitations. In connection with fractional quantum Hall liquids, the MERA network for anyons has been examined in an algebraic manner~\cite{Pfeifer,Pfeifer2,Feiguin}. Our theory uses completely different topological properties. As for topological aspects, it has been well known that the topological entanglement entropy plays dominant roles on various problems in recent condensed matter physics~\cite{Kitaev,Levin}. Thus, the complementary study from holographic viewpoints will give us very important insights to those problems.

Recently, MERA is recognized to be bulk-edge correspondence similar to the AdS/CFT correspondence in superstring theory~\cite{Maldacena}. In this context, a quantity that characterises non-locality in the original quantum system located at the boundary of the AdS space would be converted to a 'local' topological object or a topologically invariant index in the holographic space. This situation is quite similar to that of the Chern-Simons (CS) theory for topologically ordered states. Therefore it is important to find the index in our holographic space. Although we need to introduce the AdS metric in order to have some relationship between MERA and AdS/CFT, we will find that the topology affects the geometrical relationship very strongly.

\subsection{String Order Parameter and its Possible Holographic Dual}

Importance of the topological view is closely related to usefulness of hidden order parameters such as string order parameters that sometimes appear when we would like to change statistics of quantum particles. In the case of the slave particle approach to the $t$-$J$ model, the phase string recovers the edge singularity of the spinon excitation and also zero quasiparticle spectral weight~\cite{Weng,Suzuura}. Both of them can not be described by the slave boson mean field approximation. The string acts as a kind of gauge field connecting spinon and holon excitations. The presence of the gauge field reminds us a role of Yang-Mills theory on AdS/CFT. We expect that the phase string influences the topology of our holographic space.

Here, we briefly summarize the approach for later convenience. We rewrite the original electron operator as a product of spinon and holon operators as
\begin{eqnarray}
\xi_{i,\sigma}=s_{i,\sigma}h_{i}^{\dagger},
\end{eqnarray}
where both of the spinon $s_{i,\sigma}$ and the holon $h_{i}$ can be regarded as a free boson or fermion with a definite energy-momentum relation. This representation is possible when we impose the local constraint:
\begin{eqnarray}
s_{i,\uparrow}^{\dagger}s_{i,\uparrow}+s_{i,\downarrow}^{\dagger}s_{i,\downarrow}+h_{i}^{\dagger}h_{i}=1, 
\end{eqnarray}
which means that the local constraint given by inequality $\xi_{i,\uparrow}^{\dagger}\xi_{i,\uparrow} + \xi_{i,\downarrow}^{\dagger}\xi_{i,\downarrow} \le 1$ is transformed into the equation.

\begin{figure}[htbp]
\begin{center}
\includegraphics[width=5.5cm]{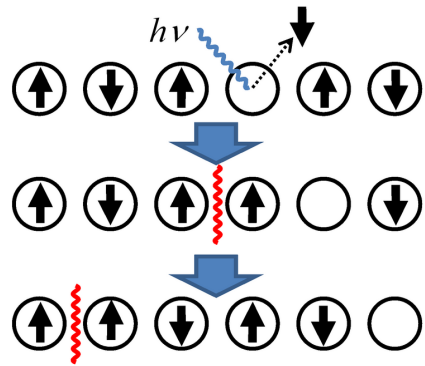}
\end{center}
\caption{Schematic viewgraph of photoemission process into the 1D Mott insulator: a wavy line represents spinon excitation, and a while circle represents holon excitation. They do not exchange their positions because of the local constraint. This means that the branching networks do not get entangled.}
\label{fig1}
\end{figure}

Then, we apply mean-field decoupling to the Hamiltonian so that the spinon and the holon parts do not interact with each other. However, this process loses the edge singularity of the spinon excitation. This is because spin and charge excitations are originally collective ones, but the approximation decomposes an electron into these excitations in the local basis. Rigolously speaking, the real collective modes associated with spin and charge degrees of freedom in bosonization and Bethe ansatz are somehow different from the present spinon and holon excitations. Thus, it may sound strange that the phase string pastes spinon and holon together and recovers the edge singularity of the spinon excitation slectrum. We should be careful for this fact in the holographic representation.

In general, the unitary transformation of $\xi_{i,\sigma}$ induces an additional phase factor represented by
\begin{eqnarray}
\xi_{i,\sigma} \rightarrow s_{i,\sigma}h_{i}^{\dagger}e^{i(\theta_{i}^{h}+\theta_{i}^{s})} , \label{sb2}
\end{eqnarray}
with 
\begin{eqnarray}
\theta_{i}^{h} &=& \mp\frac{\pi}{2}\sum_{l>i}h_{l}^{\dagger}h_{l} , \\
\theta_{i}^{s} &=& \pm\frac{\pi}{2}\sum_{l>i}\bigl(s_{l,\uparrow}^{\dagger}s_{l,\uparrow}+s_{l,\downarrow}^{\dagger}s_{l,\downarrow}-1\bigr).
\end{eqnarray}
Then, the phase factor attached to Eq.~(\ref{sb2}) is given by
\begin{eqnarray}
\theta_{i}^{h}+\theta_{i}^{s}=\mp\pi\sum_{l>i}h_{l}^{\dagger}h_{l}. \label{phase}
\end{eqnarray}
The spinon and the holon excitations always feel this extra phase. Let us consider a situation that one spinon-holon pair is created into the Mott insulator by emitting one electron at site $i$ outside of the system (see Fig.~\ref{fig1}). Then, they propagates into opposite directions. When the holon goes to the right side of $i$, the phase becomes $\pi$, and the spinon (or holon) gets the phase factor.

The holographic bifurcation in MERA is indeed a similar process to the above method, since the bifurcation is done in real space. At the same time, the system is renormalized and goes to lower and lower energy states. Then, we expect that the IR region would have important information about the phase string. Actually, the network bifurcation occurs at the IR region~\cite{Evenbly}. We will derive a holographic description of the phase string from the topology of the branching MERA network. It would be necessary to take multiple branches in many spinon-holon cases. We will also discuss the effect of multiple branches on the topological index.

\subsection{Organization of This Paper}

The paper is organized as follows. We first set up theoretical tools in Section II. Section III is the main part of this paper. We examine topology and geometry of our finite temperature MERA network in spatially 1D cases, and give holographic description of the phase string. We also examine the stability of the branching network, and answer why the branch spontaneously occurs. In Sec. IV, we consider 2D cases, and point out dimensionality dependence of the topological structure of the holographic space. In Sec. V, we discuss implications of the present results and summarize our study.

\section{Theoretical Setup}

\begin{figure}[htbp]
\begin{center}
\includegraphics[width=7cm]{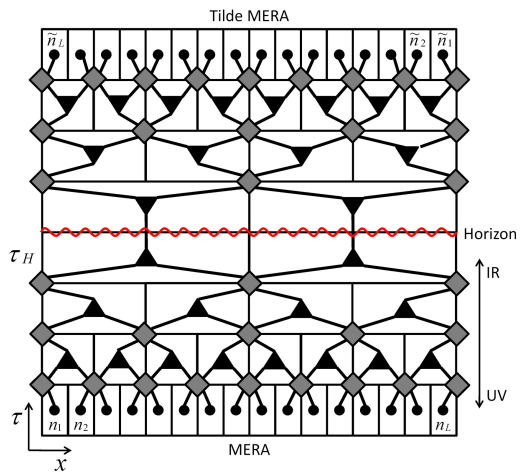}
\end{center}
\caption{MERA - tilde MERA combined network at finite temperatures. The lower half is the MERA network, and the upper half is the tilde MERA network. Filled dots, diamonds, and filled triangles are the original sites in a quantum 1D critical system ($x$ direction), disentangler tensors, and isometries, respectivery. The vertical direction denotes renormalization flow. A red wavy line represents the event horizon of the AdS black hole.}
\label{fig2}
\end{figure}

The topological study has been efficiently done at finite temperatures. The finite-temperature formalism of the MERA network has been examined in our recent paper~\cite{Matsueda,Matsueda2}. We have introduced the tilde tensor network as well as the network in the original Hilbert space by means of thermofield dynamics~\cite{Takahashi,Israel,Maldacena2,Cantcheff,Czech,Raamsdonk}. The total variational function is defined by
\begin{eqnarray}
\left|\Psi\right> = \sum_{\{n_{j}\}}C^{n_{1}\cdots n_{L};\tilde{n}_{1}\cdots\tilde{n}_{L}}\left|n_{1}\cdots n_{L};\tilde{n}_{1}\cdots\tilde{n}_{L}\right> ,
\end{eqnarray}
and
\begin{eqnarray}
C^{n_{1}\cdots n_{L},\tilde{n}_{1}\cdots\tilde{n}_{L}} = \sum_{\sigma=1}^{\chi}C_{\sigma}^{n_{1}\cdots n_{L}}C_{\sigma}^{\tilde{n}_{1}\cdots\tilde{n}_{L}} ,
\end{eqnarray}
where $C_{\sigma}^{n_{1}n_{2}\cdots n_{L}}$ is the original tensor network and $C_{\sigma}^{\tilde{n}_{1}\tilde{n}_{2}\cdots\tilde{n}_{L}}$ is its tilde conjugate. They are connected by the index $\sigma$. When they are decomposed into MERA-based tensor networks (truncated MERA network), it has been shown that the degree of freedom characterized by the index $\sigma$ represents an AdS black hole~\cite{Matsueda2,Azeyanagi,Swingle,Swingle2,Javier,Javier2}. We show a schematic viewgraph of the combined network $C^{n_{1}\cdots n_{L};\tilde{n}_{1}\cdots\tilde{n}_{L}}$ in Fig.~\ref{fig1}. Hereafter, we name the combined network as a topological manifold $X$. Because of the following technical reasons, we assume rotational invariance of $X$ (see the positions of indices $\tilde{n}_{1},\tilde{n}_{2},...,\tilde{n}_{L}$ in Fig.~\ref{fig2}). Mathematically, the assumption means
\begin{eqnarray}
\tilde{i}=L-i+1 , \label{rotation}
\end{eqnarray}
for all $i$ and for all effective site indices in renormalized layers. In the absence of the branching networks, this is simple rotation associated with symmetry between the original and the tilde Hilbert spaces. However, in the presence of the two branches, the branching position does not obey the rotational symmetry, but the mirror symmetry with respect to the event horizon.

In the continuous limit, the metric of $C_{\sigma}^{n_{1}n_{2}\cdots n_{L}}$ (or $C_{\sigma}^{\tilde{n}_{1}\tilde{n}_{2}\cdots\tilde{n}_{L}}$) is given by
\begin{eqnarray}
ds^{2}=\frac{1}{f(\tau)}d\tau^{2}+\bigl(e^{-\tau}dx\bigr)^{2} .
\end{eqnarray}
Here $f(\tau)$ is a proper function that becomes zero for $\tau\rightarrow\tau_{H}$, where the network is terminated at $\tau=\tau_{H}$. When we introduce the following coodinate transformation
\begin{eqnarray}
z=e^{\tau},
\end{eqnarray}
our result is equivalent to the AdS/CFT result with use of the metric (note that $t$ is fixed in the MERA network):
\begin{eqnarray}
ds^{2}=\frac{1}{z^{2}}\left(-f(z)dt^{2}+\frac{1}{f(z)}dz^{2}+dx^{2}\right) ,
\end{eqnarray}
where $f(z)$ and $f(\tau)$ are given by
\begin{eqnarray}
f(z) &=& 1-\frac{z^{2}}{z_{H}^{2}} , \\
f(\tau) &=& 1-e^{2(\tau-\tau_{H})} , 
\end{eqnarray}
and $z_{H}=e^{\tau_{H}}$. The layer number $\tau_{H}$ of the network corresponds to the temperature of this system
\begin{eqnarray}
\pi T=e^{-\tau_{H}}=z_{H}^{-1} .
\end{eqnarray}
Our motivations are to make $X$ in the presence of the branching structures, and to classify the topologically different structures of $X$. We also see that the shape of the black hole is important to understand the stability of the spin-charge separation in the holographic space.

\section{Topology and Geometry of 2D Holographic Space}

\subsection{Homeomorphic Transformation of the Holographic Space into a M$\ddot{\rm o}$bius Strip}

\begin{figure}[htbp]
\begin{center}
\includegraphics[width=7cm]{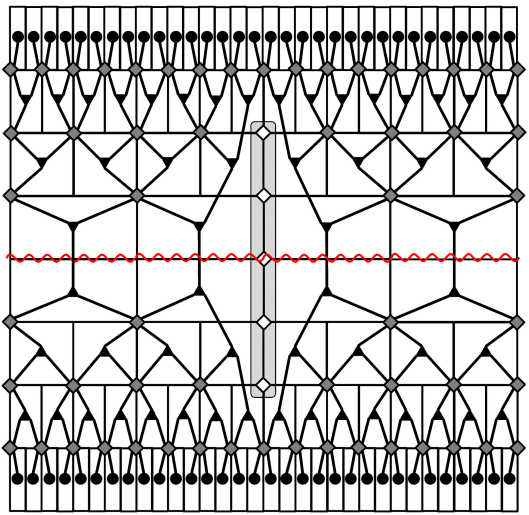}
\end{center}
\caption{Connection of two branching MERA networks. The shaded area represents a defect created in this manifold.}
\label{fig3}
\end{figure}

\begin{figure}[htbp]
\begin{center}
\includegraphics[width=8.5cm]{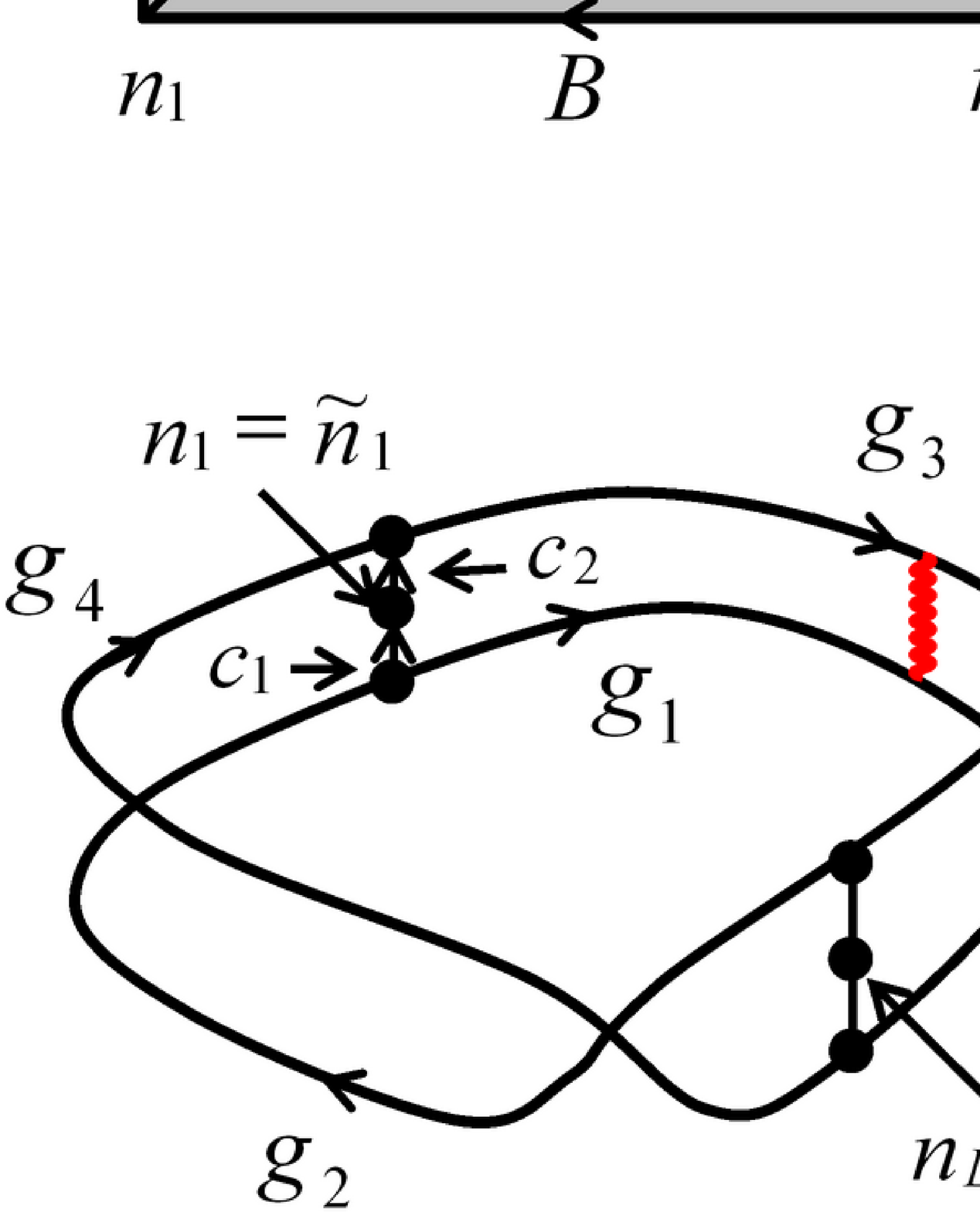}
\end{center}
\caption{Homeomorphic transformation of genus-$1$ network into a M$\ddot{\rm o}$bius strip. We take boundary conditions $T=B$ (pairing of a basis and its ancilla) and $L=R$ (twist boundary condition in $X$).}
\label{fig4}
\end{figure}

Let us first consider 1D cases. For the moment, we assume that $X$ has two branches representing a pair of spinon-holon excitations. Again note that they are spatially separated in the slave particle formalism, and different from real corrective excitations. At finite temperatures, the modification in IR region of $X$ (near the black hole horizon) necessarily induces a 'defect' in this manifold as shown in Fig.~\ref{fig3}. Due to the presence of the defect, the manifold $X$ is topologically equivalent to a M$\ddot{\rm o}$bius strip of Euler characterisric $\chi(X)=0$, if we could couple $\left|n_{i}\right>$ with $\left|\tilde{n}_{i}\right>$ for all $i$ and impose a twisted boundary condition for the left and the right edges of our finite-size system. The former condition is natural, since they are maximally entangled. Now, we consider the two branches that represents spinon and holon excitations. Then, we can not introduce a simple periodic boundary condition along $x$ direction so that they are not connected directly. As for the latter condition, the spinon $s$ (holon $h$) couples with its ancilla $\tilde{s}$ ($\tilde{h}$). We can see the coupling through the centerline on the strip,

A method for homeomorphic (topologically invariant) transformation of our manifold into the strip is shown in Fig.~\ref{fig4}. Let us define the number of the defects as $N_{g}$ in general. We also define
\begin{eqnarray}
N_{g}=2k-1, \label{ng}
\end{eqnarray}
for $k=1,2,...$, since $k$ represents the number of spinon-holon pairs. Then, $X$ has $k$-genera [Euler characteristic $\chi(X)=2-2k$], and we call $X$ as genus-$k$ network or genus-$k$ manifold. Figure~\ref{fig3} is a genus-$1$ network. In this case, the defect is transformed into the genus. Later we will give more conclete formulation. In addition, the M$\ddot{\rm o}$bius strip has a single boundary representing a close loop covering the genus edge. Starting from a particular boundary point, we go back to the original position by two loops. Thus, we get a phase $\pi$ per loop. This phase is closely related to Eq.~(\ref{phase}), and in the next section we consider the mechanism of the duality.

The phase string in the slave particle formalism is an 'extended' object because it connects a spinon with a holon moving into opposite directions. However, the genus is in some sense 'local' (and static) before taking the twisted boundary conditions, since the inside of the genus does not contain isometry tensors. This means that the area of the genus should be infinitesimally small when we introduce a proper metric near the genus. This fact is also consistent with duality requirement.

A very simple argument is that $N_{g}$ characterizes the number of the excitation modes at low temperatures. However, this is an important fact. When the genus varnishes with increasing temperature and the spinon and the holon branches are smeared out by temperature fluctuation, the system undergoes the topologically different state [$N_{g}=1 \rightarrow N_{g}=0$ (Fig.~\ref{fig2})]. According to bulk-edge correspondence in topological systems, an evidence of this 'hidden' topological transition would be detected by the topological entanglement entropy at the boundary of our AdS space. Unfortunately, it would be hard to do large-scale numerical simulation for the study of temperature dependence of the entanglement entropy in the $t$-$J$ model. In the present stage, we do not have numerical evidence of the topological transition in specific models.

If we cut the strip along the horizon, the system becomes a long strip and the twist is released. Here we suppose that the cut generates the ground state. Only that point seems to be singular in our representation of the variational function. In the original formulation of the MERA network in the ground state, the lowest energy layer is capped by the top tensor, and that means separation of $X$ into two independent MERA networks. In the present case, the top tensor still has an additional index that connects with the ancilla space in order to keep the $\pi$ phase. This indicates that the proper variational function depends on the number of the excitation modes in the original quantum systems.

\subsection{Duality between Non-Local Phase String and Genus (or Twist of M$\ddot{\rm o}$bius Strip)}

Let us formulate the correspondence between the phase string and the $\pi$ phase of M$\ddot{\rm o}$bius strip more precisely. Up to now, we have considered a particular combined-MERA network with a genus, and for instance Fig.~\ref{fig2} was not translationally invariant. In order to recover the invariance, we linearly combine a set of branching tensors with different positions~\cite{Ostlund,Ueda}.

When the position of the defect at the critical layer $\tau=\tau_{c}$ is labeled by $p$, the translationally invariant anzats is represented as 
\begin{eqnarray}
\left|\Psi\right> &=& \sum_{p}T_{p}\left|\Psi_{p}\right> , \\
\left|\Psi_{p}\right> &=& \sum_{\{n_{j}\}}\sum_{\sigma=1}^{\chi}C_{\sigma,p}^{n_{1}\cdots n_{L}}C_{\sigma,N-p}^{\tilde{n}_{1}\cdots\tilde{n}_{L}} \nonumber \\
&& \times\left|n_{1}\cdots n_{L}\right>\left|\tilde{n}_{1}\cdots\tilde{n}_{L}\right> ,
\end{eqnarray}
where $N=2^{-\tau_{c}}L$. We again note that we consider the twisted boundary condition for the $x$ dierection. The tensor $C_{\sigma,p}^{n_{1}\cdots n_{L}}$ is decomposed into bottom tensor $B$, left-branch tensor $R$, and right-branch tensor $L$:
\begin{eqnarray}
C_{\sigma,p}^{n_{1}\cdots n_{L}} = \sum_{\{m_{j}\}}B^{n_{1}\cdots n_{L}}_{m_{1}m_{2}\cdots m_{N}}L_{\sigma_{1}}^{m_{1}\cdots m_{p}}R_{\sigma_{2}}^{m_{p+1}\cdots m_{N}} ,
\end{eqnarray}
where $\sigma=(\sigma_{1},\sigma_{2})$. We proceed decomposition as follows:
\begin{eqnarray}
&& \sum_{\{n_{j}\}}C_{\sigma,p}^{n_{1}\cdots n_{L}}\left|n_{1}n_{2}\cdots n_{L}\right> \\
&& = \sum_{\{n_{j}\}}\sum_{\{m_{j}\}}B^{n_{1}\cdots n_{L}}_{m_{1}\cdots m_{N}}L_{\sigma_{1}}^{m_{1}\cdots m_{p}}R_{\sigma_{2}}^{m_{p+1}\cdots m_{N}} \nonumber \\
&& \;\;\;\;\;\;\;\;\;\;\;\;\;\;\;\;\;\;\times\left|n_{1}\cdots n_{L}\right> \\
&& = \sum_{\{m_{j}\}}L_{\sigma_{1}}^{m_{1}\cdots m_{p}}R_{\sigma_{2}}^{m_{p+1}\cdots m_{N}}\Phi_{m_{1}\cdots m_{N}} ,
\end{eqnarray}
with
\begin{eqnarray}
\sum_{\{n_{j}\}}B^{n_{1}\cdots n_{L}}_{m_{1}\cdots m_{N}}\left|n_{1}\cdots n_{L}\right> = \Phi_{m_{1}\cdots m_{N}} .
\end{eqnarray}
Finally, the total wave function is given by
\begin{eqnarray}
\left|\Psi\right> &=& \sum_{\{m_{j}\}}X^{m_{1}\cdots m_{N}}_{\tilde{m}_{1}\cdots\tilde{m}_{N}}\Phi_{m_{1}\cdots m_{N}}\Phi_{\tilde{m}_{1}\cdots\tilde{m}_{N}} ,
\end{eqnarray}
with
\begin{eqnarray}
X^{m_{1}\cdots m_{N}}_{\tilde{m}_{1}\cdots\tilde{m}_{N}} &=& \sum_{p}T_{p}\sum_{\sigma_{1}}(L_{\sigma_{1}}^{m_{1}\cdots m_{p}}R_{\sigma_{1}}^{\tilde{m}_{N-p+1}\cdots\tilde{m}_{N}}) \nonumber \\
&& \times\sum_{\sigma_{2}}(L_{\sigma_{2}}^{\tilde{m}_{1}\cdots\tilde{m}_{N-p}}R_{\sigma_{2}}^{m_{p+1}\cdots m_{N}}) .
\end{eqnarray}
It is noted that we should be careful for the indices of $L$ and $R$.

\begin{figure}[htbp]
\begin{center}
\includegraphics[width=7.5cm]{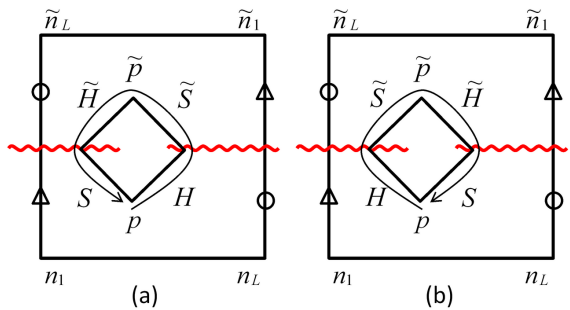}
\end{center}
\caption{Spinon and holon excitations in the holographic space (single electron case). (a) $L=S$ (spinon) and $R=H$ (holon), and (b) $L=H$ and $R=S$. Circles and triangles denote twisted boundary condition along $x$ direction.}
\label{fig5}
\end{figure}

The tensors $L$ and $R$ for $\tau\ge\tau_{c}$ represent the spinon and the holon branches, respectively. The tensor operation to $\Phi$ changes the basis state. One of them represents the holon creation, and the other annihilates the spinon excitation, since we are now considering one-electron emitted state $\xi_{i,\sigma}=s_{i,\sigma}h_{i}^{\dagger}$. For one-electron excitation process at site $i$ ($\rightarrow p$ after renormalization), there are two patterns of the spinon-holon configurations: one is that the spinon is created at the left hand side of $i$ [Fig.~\ref{fig5}(a)], while the other one is that the holon is created at the left hand side of $i$ [Fig.~\ref{fig5}(b)]. Because of the presence of the tilde space, the one-particle excitation process is always follwed by its tilde conjugate process. Therefore, the process forms a close loop along the genus as shown in Fig.~\ref{fig5}. Figures.~\ref{fig5}(a) and (b) can be classified by the direction of the rotation along the genus. We put the direction on each figure so that one-electron emission process $\xi_{i,\sigma}=s_{i,\sigma}h_{i}^{\dagger}$ can be imagined. Since our system is a twisted strip, the excitation $sh^{\dagger}$ (or $\tilde{s}\tilde{h}^{\dagger}$) representing half loop along the genus gets the $\pi$ phase of the M$\ddot{\rm o}$bius strip. This is what we have already observed in Eq.~(\ref{phase}). 

These processes are also characterised by fibre bundle $(X,\pi,E,F,G)$. Let us identify $X$ to be a fibre space, $E$ to be a base space ($S^{1}$) representing the edge of the genus ($\pi:E\rightarrow X$), and $F$ to be a fibre. We also introduce the structure group $G$ that is the group of homeomorphism of $F$. Then, the structure group $G$ is equivalent to ${\bf Z}_{2}=\{e,g\}$ where $e$ is identity map and $g$ denotes inversion of the fibre. This ${\bf Z}_{2}$ symmetry produces the $\pi$ phase.

Let us assume that the holographic space is completely classical. We have known that the tensor lank at the horizon is given by $e^{c/6}$, and this is close to $1$ for the Gaussian CFT ($c=1$). Thus, this assumption is reasonable near the genus. Then, $\pi/2$-rotation induces excitations such as $s^{\dagger}\tilde{h}$. These excitations are combined strongly by hidden quantum entanglement induced by the horizon, since the horizon is generated as a result of maximally entanglement. This is also one representation of the phase string in this coodinate.

\subsection{Induced Metric of a M$\ddot{\rm o}$bius Strip}

Let us look at geometrical properties of the strip. A parametric representation of the M$\ddot{\rm o}$bius strip embedded into Euclidean space ${\bf R}^{3}$ is given by
\begin{eqnarray}
x &=& \cos 2\theta(r\cos\theta + 2) , \\
y &=& \sin 2\theta(r\cos\theta + 2) , \\
z &=& r\sin\theta ,
\end{eqnarray}
where the parameters $r$ and $\theta$ are taken to be $-1\le r\le 1$ and $0\le \theta\le\pi$. Here, the center line of M$\ddot{\rm o}$bius strip is located at $r=0$, and the both edges of the strip are given by $r=\pm 1$. The center line is a circle with radius $2$. For $\vec{p}=(x,y,z)$ and the renamed coodinate $(u_{1},u_{2})=(r,\theta)$, the induced metric is defined by
\begin{eqnarray}
g_{\mu\nu}=\left<\frac{\partial\vec{p}}{\partial u_{\mu}} , \frac{\partial\vec{p}}{\partial u_{\nu}}\right> ,
\end{eqnarray}
where the bracket denotes the inner product. We obtain
\begin{eqnarray}
g_{rr} &=& 1 , \\
g_{r\theta} &=& 0 , \\
g_{\theta\theta} &=& 4\bigl(r\cos\theta+ 2\bigr)^{2}+r^{2}.
\end{eqnarray}
Therefore, the square of the line element is given by
\begin{eqnarray}
ds^{2} = dr^{2} + \left\{4(r\cos\theta + 2)^{2}+r^{2}\right\}d\theta^{2}. \label{metric}
\end{eqnarray}
When we introduce the polar coodinate $x=r\cos\theta$ and $y=r\sin\theta$, we obtain
\begin{eqnarray}
ds^{2} = dx^{2} + dy^{2} + \frac{4(x+2)^{2}}{(x^{2}+y^{2})^{2}}(-ydx+xdy)^{2} . \label{metric2}
\end{eqnarray}

Unfortunately, the mapping from the M$\ddot{\rm o}$bius strip into $X$ is not isometric. This is because the metric of $X$ should be hyperbolic at least for the main framework of $L$ and $R$. However, Eq.~(\ref{metric}) still contains enough topological information of $X$. We clearly see that the edges labeled by $r=\pm 1$ represent the left and the right branches, $L$ and $R$. For a fixed $\theta$ value, the metric becomes asymmetric with respect to the sign change of $r$ because of the presence of the factor $2$ in $(r\cos\theta + 2)^{2}$. This factor comes from the tortion of the strip. When we change the sign of $r$, the line element remains unchanged under the additional sign change $\theta\rightarrow -\theta$.

Phenomenologically, it may be possible to introduce an effective metric that is obtained by deviding the right hand side of Eq.~(\ref{metric2}) by $y^{2}$. The UV region of this new metric does not approach asymptotically AdS, but our attention is to look at the IR behavior that represents two CFTs for spinon and holon excitations. In this context, that would not be bad representation.

\subsection{Effect of Multiple Branches on the Phase String}

For $N_{g}=2$, we have three branches in the holographic space, and then one extra branch appears. This is not a physical situation, and we can not introduce any physical boundary condition. In order to consider multiple spinon-holon pairs, we need to take odd number of the genus number $N_{g}$.

\begin{figure}[htbp]
\begin{center}
\includegraphics[width=8cm]{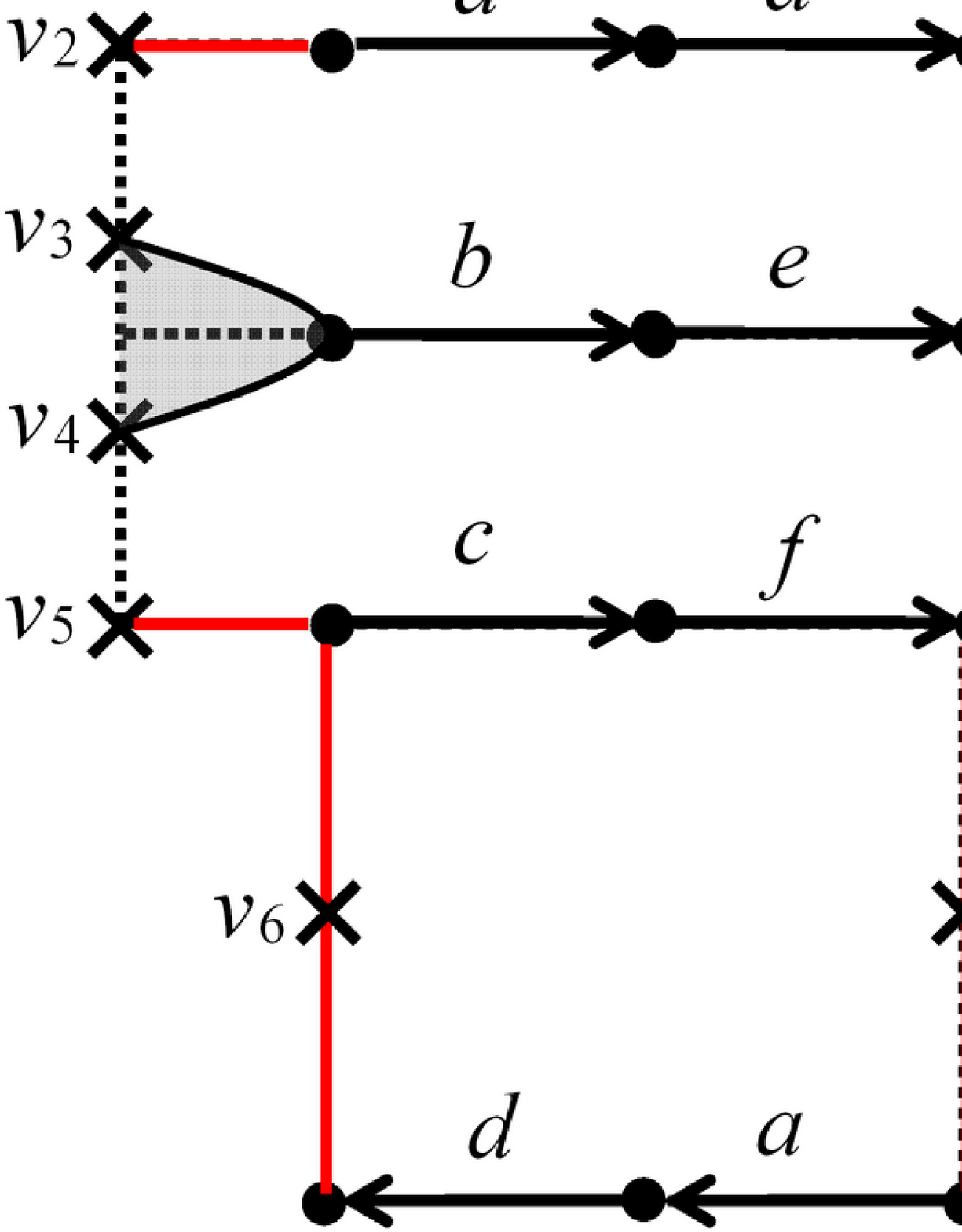}
\end{center}
\caption{Homeomorphic transformation of genus-$2$ network. We assume twisted boundary condition along $\delta$.}
\label{fig6}
\end{figure}

\begin{figure}[htbp]
\begin{center}
\includegraphics[width=5cm]{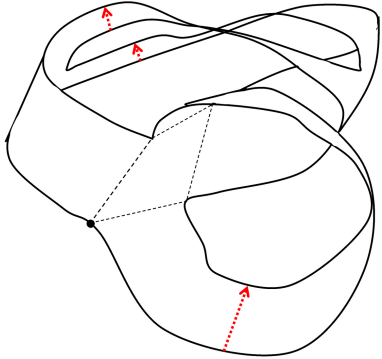}
\end{center}
\caption{Two entangled M$\ddot{\rm o}$bius strips. Red dashed arrows are the cutting positions for counting $p_{1}(X)$.}
\label{fig7}
\end{figure}

In cases of $N_{g}=3$ ($k=2$), there are two patterns of the spinon-holon configuration in $X$: one is $S-H-S-H$, and the other is $S-H-H-S$. The former case requires the twisted boundary condition along the $x$ direction, while in the latter case we can take the normal periodic boundary condition. Figure~\ref{fig6} show a method for homeomorphism in the case of $S-H-S-H$. The manifold $X$ is transformed into two entangled M$\ddot{\rm o}$bius strips as shown in Fig.~\ref{fig7}, and one of them has a 'worm hole' arising from the center defect before taking the boundary conditions. The boundary of $X$ forms a closed path, and then the left and the right defects play a role on the twist of this space.

Here we characterize this structure by topological indices. The Euler characteristic $\chi(X)$ is given by
\begin{eqnarray}
\chi(X)=p_{0}(X)-p_{1}(X)+p_{2}(X) ,
\end{eqnarray}
where $p_{\mu}(X)$ ($\mu=0,1,2$) are the $\mu$-th Betti numbers. We find that $p_{0}(X)=1$, $p_{1}(X)=N_{g}$, and $p_{2}(X)=0$. Thus, we obtain
\begin{eqnarray}
\chi(X)=1-N_{g}. \label{Euler}
\end{eqnarray}
In the present case, we have $\chi(X)=-2$. We have confirmed that the value is equivalent to that obtained from the Euler-Poincar$\acute{\rm e}$ formula under triangular decomposition of $X$. Substituting Eq.~(\ref{ng}) into Eq.~(\ref{Euler}), we obtain
\begin{eqnarray}
\chi(X)=2-2k,
\end{eqnarray}
which is nothing but the definition of the genus number $k$. The global structure of our manifold is the same as that of the torus with $2$ genera. Thus, we know that two M$\ddot{\rm o}$bius structures in Fig.~\ref{fig7} are essential for topological properties of this system. Actually, the worm hole does not twist the space.

Let us consider a closed path on the boundaries of $X$. Our starting point is the entangled position between them marked by a filled dot in Fig.~\ref{fig7}. First, we go along half of the boundary of one strip, and the course suddenly jumps into the boundary of the other strip. We further go along half of the boundary, and then we go back to the original position. Totally, we pick up the phase $2\pi$. We can interprete the present result as a definition of string in the holographic space that the $\pi$ phase winds itself round a genus which is located in between spinon (left) and holon (right) branches. In general, a close loop induces a phase factor $k\pi$. This is consistent with Eq.~(\ref{phase}).

\subsection{Topology of AdS Black Holes and Stability of Spin-Charge Separation}

When we impose the AdS metric near the genus, we find that geometric study of our manifold immediately leads to some important aspects associated with the second law of black hole thermodynamics. We look at geometric properties near the genus, and discuss why the spin charge-separation is stable in 1D quantum systems. In the following, we consider a case of $N_{g}=1$, but the discussion can be extended to more complicated cases.

\begin{figure}[htbp]
\begin{center}
\includegraphics[width=8cm]{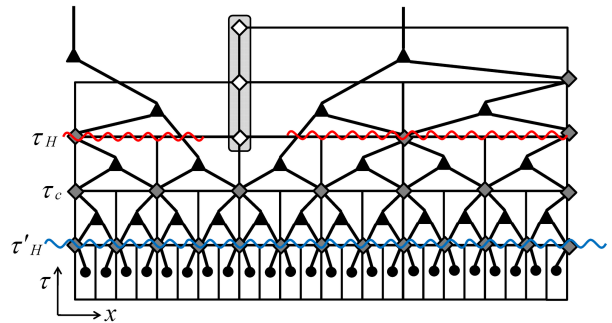}
\end{center}
\caption{Temperature dependence of the total area of the AdS black holes.}
\label{fig8}
\end{figure}

The second law states that the total area of the black hole should be larger than the sum of each black hole area, when more than two black holes merge into single one through collapsing processes. In our case, the collapse occurs when we increase temperature through a critical value $\tau_{c}$. Let us look at Fig.~\ref{fig8}. We take $A$ as the number of the connecting bonds between the MERA and the tilde MERA (this quantity is equal to the black hole area). For $\tau_{H}>\tau_{c}>\tau_{H}^{\prime}$, we clearly observe
\begin{eqnarray}
A(\tau_{H}^{\prime},N_{g}=0) > A_{L}(\tau_{H},N_{g}=1) + A_{R}(\tau_{H},N_{g}=1) , \label{inequality}
\end{eqnarray}
where $A_{L}$ and $A_{R}$ are left and right branches, respectively. Here, the black hole entropy is defined by
\begin{eqnarray}
S_{BH}=\ln\chi=A\ln m ,
\end{eqnarray}
where $\chi=m^{A}$ is total degree of freedom at the horizon and $m$ is the tensor dimension of the isometry at the horizon ($m=e^{c/6}$)~\cite{Matsueda2}. Thus, we find that Eq.~(\ref{inequality}) is nothing but the second law.

As already mentioned, the inside of the genus only contains disentangler tensors. We find (for the same $L$)
\begin{eqnarray}
A(\tau_{H},N_{g})=A(\tau_{H},N_{g}^{\prime}), \label{balance}
\end{eqnarray}
and thus we obtain $S_{BH}(\tau_{H},N_{g})=S_{BH}(\tau_{H},N_{g}^{\prime})$. This result satisfies the necessary condition $S(\tau_{H},N_{g})\le S(\tau_{H},N_{g}^{\prime})$ for $N_{g}>N_{g}^{\prime}$. Equation~(\ref{balance}) indicates that in a view point of entropy counting the probability of a single black hole is equal to that of creation of some black holes. Then, spontaneous separation of $X$ due to this equality is a possible scenario. After the branch, left and right branches can not contact with each other, and this leads to the spin-charge separation.

\subsection{Comparison with Quantum Monte Carlo Calculation}

The temperature dependence of the spinon and the holon excitations has been numerically examined by the quantum Monte Carlo simulation~\cite{Matsueda3}. In this examination, the single-particle spectral function $A(\vec{k},\omega)$ at $T=0.25t$ was compared with the dynamical density matrix rebormalization group result for zero temperature. Then, we have observed that the spinon and the holon singularities are still visible even such a high temperature. Actually, an evidence of these singularities has been observed in angle-resolved photoemission spectroscopy in 1D cuprates such as SrCuO${}_{2}$ and Sr${}_{2}$CuO${}_{3}$~\cite{Kim}. These data indicate that the value of $\tau_{c}$ is small. In this case, an additional entropy is expected to be small.

\section{Topology of 3D Holographic Space}

\begin{figure}[htbp]
\begin{center}
\includegraphics[width=7.5cm]{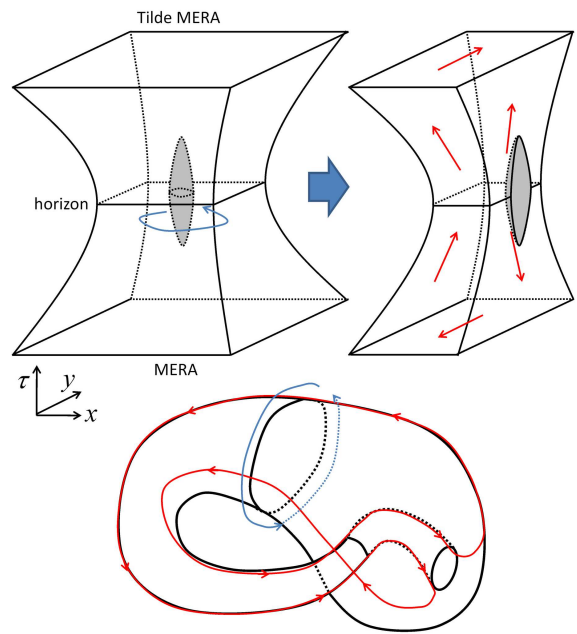}
\end{center}
\caption{A defect in the 3D holographic space. This manifold is converted into a Klein bottle under homeomorphic transformation. A red closed route on the bottle is equal to the boundary of a M$\ddot{\rm o}$bius strip, and thus the route devides the bottle into two strips. Note that there is no self-crossing if we could draw the a figure of the bottle in ${\bf R}^{4}$.}
\label{fig9}
\end{figure}

Let us move to 2D quantum critical systems. We present a schematic viewgraph of the 3D holographic space in Fig.~\ref{fig9} (unfortunately it is complicated to represent our discrete space by the tensor network). We have introduced a line defect in IR region of our network. This is also composed of only disentangler tensors. After homeomorphic transformation, we know that the outside area of the defect is equivalent to the surface of a Klein bottle with $\chi(X)=0$ and $k=1$. Once we devide the manifold into two parts as shown in the right upper panel of Fig.~\ref{fig9}, and we see that the edge of the defect is equal to the boundary of the M$\ddot{\rm o}$bius strip under an appropriate boundary condition. Pasting two M$\ddot{\rm o}$bius strips together leads to a Klein bottle as shown in the bottom panel of Fig.~\ref{fig9} (in the present case the inside of the bottle is occupied by the tensor network).

At first, we see that completely separated branches do not exist. For instance, it is possible to consider an rotating excitation shown by an allows. The system is invariant under rotational transformation around the defect in the continuous limit. Then, we can not distinguish one excitation with another. In the Klein bottle, such a feature is represented by a close loop along the surface. These features well explain the strong mixing of spin and charge excitations in 2D.

An equation of motion approach for the $t$-$J$ model shows that the composite electrons dressed with charge and spin fluctuation strongly couple with each other particularly near the wave vector $\vec{k}=(\pi/2,\pi/2)$ for nearly optimum hole doping where the presence of the quantum critical point is expected~\cite{Matsueda4}. As is well known, the coupling and the separation among various quantum degrees of freedom depend on spatial dimensionality.

In viewpoints of exotic electronic states realized in 2D, such as high-$T_{c}$ superconductivity, we would like to determine the metric of this system and calculate physical quantities. Recently, AdS/CFT has been applied to the examination of such electronic states~\cite{Liu,Ogawa,Huijse,Shaghoulian}. It is important to study relatioship between them and our approach.

\section{Remarks and Summary}

To undestand the separation and the coupling of various quantum degrees of freedom are always fundamental issues in the study of strongly correlated electron systems. The present study opens a unique route to look at such a feature in the holographic space. In order to proceed theoretical examination in realistic models, it will be a first step to calculate the holographic entanglement entropy (geodesic distance), and to compare it with the topological entanglement entropy in the original quantum system. Fortunately, the metric of the branching MERA network has been obtained in Eq.~(\ref{metric}) (we need to embed it into hyperbolic space). Our future work will be such calculation straightforwardly.

Summarizing, we have examined various topological aspects associated with bulk-edge correspondence of branching MERA networks at finite temperatures. The topological view is quite important. Even if we do not introduce the metric, we can pick up fundamental electronic properties of our target models by analysing topology of our holographic space. In terms of strongly correlated electron physics, our examination is closely related to the spin-charge separation and coupling based on the slave particle formalism as well as the phase string in the $t$-$J$ model. We have found that a holographic dual of the phase string is the twist of the network along the genus created in the IR region. We have also examined the stability of the branching networks in connection with the second law of black hole thermodynamics. Then, we have proved that the MERA network for one-dimensional quantum critical systems spontaneously separates into multiple branches in the IR region of the network, while the branch does not occur in more than two dimensions. The sharp difference clearly illustrates the spin-charge separation in 1D, and the spin-charg ecoupling in 2D. Our finding, that a holographic dual of the phase string is the network topology, is realistic extension of the CS theory, and will provide us unique topological properties of extra dimension. The CS theory is a kind of bulk-edge correspondence, and is considered to be a toy model of AdS/CFT. This is because the CS theory introduces fictitious gause field and does not treat gravitation explicitly. My approach treats topological properties of our spacetime as the CS theory does, but is basically an AdS/CFT-oriented method. I hope that the present result contributes to bringing up more golobal and essential viewpoints for bulk-edge correspondence.

{\it Note added:} In recent arXiv by T. Takayanagi et at.~\cite{Nozaki}, they have briefly commented on the MERA for topologically ordered phases and its connection to D-branes~\cite{Aguado,Fujita}. There, the top tensor of MERA contains topological information such as ground-state degeneracy. In my case, the top tensor is separated into some parts, and in this case the topology around them becomes also important.

\begin{acknowledgements}
The author thanks Masafumi Ishihara for discussion, and also thanks Tadashi Takayanagi for giving some comments on the manuscript.
\end{acknowledgements}

\end{document}